\documentclass[prb,twocolumn,superscriptaddress,showpacs,amsmath,amssymb,nofootinbib]{revtex4-1}

\usepackage{graphicx}
\usepackage{latexsym}
\usepackage{amsmath}
\usepackage{amssymb}
\usepackage{amsfonts}
\usepackage{color}
\usepackage{bm}% bold math
\usepackage{verbatim}
\usepackage[]{color}
\definecolor{red}{rgb}{1,0,0}
\usepackage{framed}
\definecolor{shadecolor}{RGB}{222,222,221}
\begin{document}

\title{Reconstruction of the DOS  at the end of a S/F bilayer.}

 \date{\today}

\author{I. V. Bobkova}
\affiliation{Institute of Solid State Physics, Chernogolovka, Moscow
  reg., 142432 Russia}
%\affiliation{Moscow Institute of Physics and Technology, Dolgoprudny, 141700 %Russia}

\author{A. M. Bobkov}
\affiliation{Institute of Solid State Physics, Chernogolovka, Moscow reg., 142432 Russia}

 \begin{abstract}
Influence of a nonmagnetic surface on the density of states in a spin-split superconductor, which is realized on the basis of a S/F bilayer, is investigated. It is demonstrated that if the ferromagnet magnetization has a defect in the form of a domain wall in the vicinity of the surface, the superconducting density of states is reconstructed manifesting the spin-split Andreev resonances. Formation of these resonances is not connected to any superconducting order parameter inhomogeneities. Andreev reflection processes forming the resonances occur at the inhomogeneity of the spin-split gap generated by the domain wall. These resonances can be used as a spectroscopic probe of a domain wall presence and motion.   
 \end{abstract}

 %%% PACS numbers
  \maketitle
 
\section{introduction}
 
Active development of superconducting spintronics, caloritronics and spin caloritronics raised a renewed interest in studying hybrid structures of superconductors and ferromagnets. This research is motivated, in particular,
by the desire to employ the spin torques generated by the dissipationless spin supercurrents \cite{Grein2009,Alidoust2010,Shomali2011,Brydon2011,Moor2015,Gomperud2015,Halterman2015,Bobkova2004,Alidoust2015,Jacobsen2016,Konschelle2016,Montiel2018,Ouassou2018}. The possibility to have dissipationless spin-polarized and purely spin supercurrents is based on the equal-spin triplet proximity effect \cite{Bergeret2001, Bergeret2004, Bergeret2005b, Bergeret2005,Eschrig2015}. 

Another extremely important for superconducting spintronics property of S/F hybrids is the spin-split superconducting density of states (DOS). The last property is a key ingredient for superconducting spintronics and spin caloritronics because a series of interesting
phenomena have been studied in superconductor/ferromagnet structures
with spin-split DOS. Among them are the giant
thermoelectric \cite{Machon2013, Ozaeta2014, Kolenda2016, Kolenda2016_1, Giazotto2014, Giazotto2015}, thermospin \cite{Ozaeta2014,Linder2016,Bobkova2017} effects, highly efficient spin and heat valves \cite{Huertas-hernando2002, Giazotto2013,Giazotto2013_2,Giazotto2006,Giazotto2008}, cooling at the nanoscale \cite{Giazotto2006_2,Kawabata2013} and low-temperature thermometry and development of sensitive electron thermometers \cite{Giazotto2015_2}. Detailed discussion of the modern advances associated with the spin-split superconducting DOS can be found in the recent review\cite{Bergeret_review}.

\begin{figure}[!tbh]
   \centerline{\includegraphics[clip=true,width=3.2in]{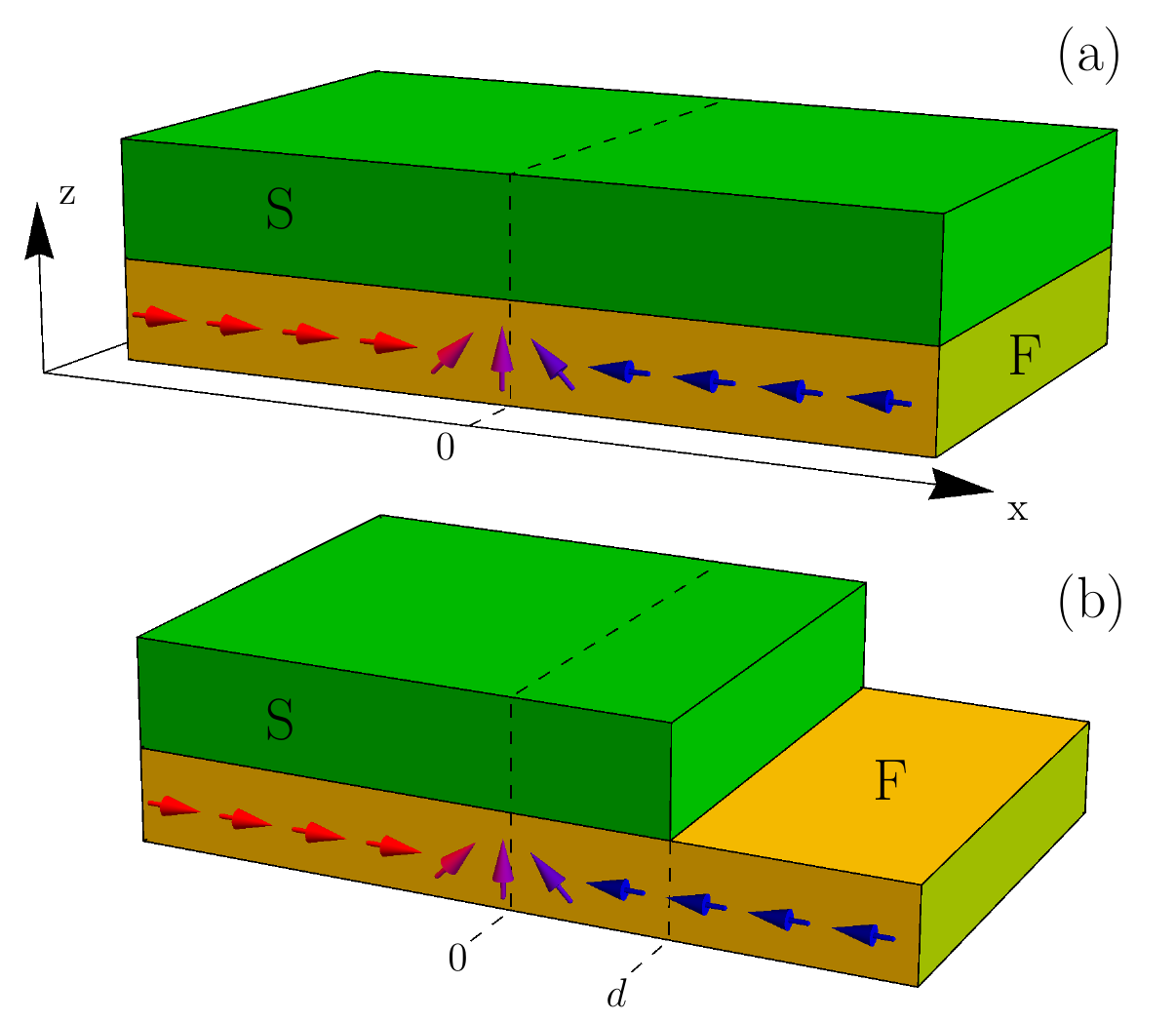}}
     \caption{(a) Infinite S/F bilayer with a domain wall; (b) S/F bilayer with a closed end of a superconductor. The DW center is at the distance $d$ from the end of the superconductor.}
 \label{sketch}
 \end{figure}

In spite of great interest in spin-split superconductors the research of the spin-split DOS in hybrids with textured ferromagnets is only at the very beginning. The influence of the domain structure on the position-averaged superconducting DOS in S/FI bilayer was studied as experimentally, so as theoretically \cite{Strambini2017}. The DOS in the ferromagnetic part of a metallic S/F bilayer with an infinitely sharp domain wall has been investigated \cite{Golubov2005}. The influence of the particular noncolinear texture of the domain walls and other inhomogeneous ferromagnets on the DOS in the infinite S/F bilayers was also investigated \cite{Bobkova_prep}. In the present work we study the reconstruction of the DOS at the closed end of the superconductor when it is a part of a S/F bilayer. The system under consideration is presented in Fig.~\ref{sketch}.

It is very well known that for a conventional $s$-wave superconductor the DOS is of the bulk BCS-like shape even in the vicinity of the surface. This is not true for surfaces of unconventional superconductors, where the superconducting order parameter can be suppressed at the surface and the Andreev bound states can appear \cite{Kashiwaya2000,Lofwander2001,Golubov2004}. That is, a surface can be pair breaking for unconventional superconductivity. It is also known that for conventional $s$-wave superconductivity a magnetic surface or interface with a magnetic insulator is pair breaking and the Andreev surface bound states can also arise at such a surface or interface \cite{Tokuyasu1988,Fogelstrom2000,Bobkova2005,Andersen2005}. 

In the present paper we investigate the influence of a conventional nonmagnetic surface on the DOS in a spin-split superconductor. It is easily seen that for a case of homogeneous spin-splitting the DOS at the surface of the closed end is not modified with respect to the bulk just like the case of a surface of a conventional $s$-wave BCS superconductor. However, this is not the case if the exchange field of the ferromagnet is spin textured. For concreteness we study the magnetic inhomogeneity in the form of a domain wall (DW). If the magnetic DW is close enough (at the distances of several coherence lengths) to the end of the bilayer, the DOS at the end is modified. The most striking feature of the DOS reconstruction is the appearance of the Andreev resonances in the region between the DW and the end of the bilayer. We unveil the mechanism  of these quasi-bound states formation and demonstrate that it does not connected to pair breaking of superconductivity by the surface. It is a direct consequence of the spatially-dependent  spin-splitting of the DOS in the system. The interest of these characteristic DOS features is manifold. They can be used as a spectroscopic probe of the DW presence and as a spectroscopic detector of the DW motion.  The DOS reconstruction also should be taken into account if one thinks about operating the device by nonequilibrium spin injection. It is also worth noting that if one is interested in the proximity-induced DOS in the {\it ferromagnetic} part of the bilayer, then the DOS can be modified due to the closed end of the superconductor even for the case of a homogeneous ferromagnet \cite{Golikova2012}.

\section{model and method}

The model system, sketched in Fig.~\ref{sketch} consists of the spin-textured ferromagnet with a spatially dependent exchange field $\bm h(\bm r)$ and a spin-singlet superconductor. It is assumed that the S film is in the ballistic limit and its thickness $d_t$ is small as compared to the superconducting coherence length $\xi_s = \Delta/v_F$. The ferromagnet can be a metal or an insulator. It is widely accepted in the literature that if $d_t \lesssim \xi_s$ the magnetic proximity effect, that is the influence of the adjacent ferromagnet on the S film can be described by adding the effective exchange field $h_{eff}(\bm r)$ \cite{Bergeret2001_2} to the quasiclassical Eilenberger or Usadel equation, which is used for treating the superconductor. While for the ferromagnetic insulators the magnetic proximity effect is not so simple and in general is not reduced to the effective exchange only \cite{Cottet2009,Eschrig2015_2}, we neglect the other terms (which can be viewed as additional magnetic impurities in the superconductor) in the framework of the present study and focus on the effect of the spin texture. The S film is described by the Eilenberger equation for the retarded Green's function:
\begin{eqnarray}
 i \bm v_F \nabla\check g(\bm r, \bm p_F)+\Bigl[ \varepsilon \tau_z  + \bm h_{eff}(\bm r) \bm \sigma \tau_z - \check \Delta,\check g \Bigr] = 0,~~~~~~
 \label{eilenberger}
\end{eqnarray}

where $\Delta$ is the effective order parameter in the film, which is reduced to some extent with respect to the bulk value due to the suppression by the proximity to the ferromagnet. Here in the ballistic case it is convenient to use the so-called Riccati parametrization for the Green's function \cite{Eschrig2000,Eschrig2009}. In terms of the Riccati parametrization the retarded Green's function takes the form:
\begin{eqnarray}
\check g =
\left(
\begin{array}{cc}
1+\hat \gamma \hat {\tilde \gamma} & 0 \\
0 & 1+\hat {\tilde \gamma} \hat \gamma \\
\end{array}
\right)^{-1}
\left(
\begin{array}{cc}
1-\hat \gamma \hat {\tilde \gamma} & 2 \hat \gamma \\
2 \hat {\tilde \gamma} & -(1-\hat {\tilde \gamma} \hat \gamma) \\
\end{array}
\right)
\label{riccati_GF}
\end{eqnarray}
where $\hat \gamma$ and $\hat {\tilde \gamma}$ are matrices in spin space. Please note that they differ from the conventional definition \cite{Eschrig2000,Eschrig2009} by factors $i\sigma_y$ as $\hat \gamma_{standard} = \hat \gamma i \sigma_y$ and $\hat {\tilde \gamma}_{standard} = i \sigma_y \hat {\tilde \gamma}$.
The Riccati parametrization Eq.~(\ref{riccati_GF}) obeys the normalization condition $\check g^2 = 1$ automatically.

The Riccati amplitudes $\hat \gamma$ and $\hat {\tilde \gamma}$ obey the following Riccati-type equations:
\begin{eqnarray}
 i \bm v_F \nabla \hat \gamma + 2 \varepsilon  \gamma = -\Delta^* \hat \gamma^2 - \bigl\{ \bm h_{eff} \bm \sigma, \hat \gamma \bigr\} - \Delta ,
 \label{riccati}
\end{eqnarray}
and $\hat {\tilde \gamma}$ obeys the same equation with the substitution $\varepsilon \to -\varepsilon$, $\bm h_{eff} \to -\bm h_{eff}$ and $\Delta \to \Delta^*$. In this work we assume $\Delta = \Delta^*$. Moreover, we neglect the spatial variations of the order parameter and assume $\Delta = const$.
If we consider a locally spatially inhomogeneous  magnetic texture like a domain wall the Riccati amplitudes $\hat \gamma$ and $\hat {\tilde \gamma}$ can be found from Eq.~(\ref{riccati}) numerically with the following asymptotic condition:
\begin{eqnarray}
 \hat \gamma_{\infty} = \gamma_{0\infty} + \frac{\bm h_{eff,\infty} \bm \sigma}{h_{eff}} \gamma_\infty ,~~~~~~~~~~~~~~~~~~ \\
 \gamma_{0\infty} = -\frac{1}{2}\Bigl[ \frac{\Delta}{\varepsilon +h_{eff}+i\sqrt{\Delta^2 - (\varepsilon + h_{eff})^2}} + \nonumber \\ \frac{\Delta}{\varepsilon-h_{eff}+i\sqrt{\Delta^2 - (\varepsilon  - h_{eff})^2}} \Bigr],~~~~~~~~ \\
 \gamma_\infty = -\frac{1}{2}\Bigl[ \frac{\Delta}{\varepsilon+h_{eff}+i\sqrt{\Delta^2 - (\varepsilon  + h_{eff})^2}} - \nonumber \\ \frac{\Delta}{\varepsilon -h_{eff}+i\sqrt{\Delta^2 - (\varepsilon  - h_{eff})^2}} \Bigr],~~~~~~~~~~~~
  \label{riccati_asymptotic}
\end{eqnarray}
and $\hat {\tilde \gamma}_\infty = - \hat \gamma_\infty$.

Eq.~(\ref{riccati}) is numerically stable if it is solved starting from $x = -\infty$ for right-going trajectories $v_x > 0$ and from $x = +\infty$ for left-going trajectories $v_x < 0$. On the contrary, $\hat {\tilde \gamma}$ should be found numerically starting from $x = +\infty$ for right-going trajectories $v_x > 0$ and from $x = -\infty$ for left-going trajectories $v_x < 0$.
Having at hand $\hat \gamma(x,\bm p_F)$ and $\hat {\tilde \gamma}(x, \bm p_F)$ it is possible to calculate the DOS as follows:
\begin{eqnarray}
N= N_F{\rm Re}\Bigl\{{\rm Tr}\langle \hat g \rangle\Bigr\}
\label{DOS_ballistic}
\end{eqnarray}
where $N_F$ is the normal state DOS per spin, $\langle ... \rangle$ means averaging over the 2D Fermi surface of the superconducting film and $\hat g = (1+ \hat \gamma \hat {\tilde \gamma})^{-1}(1 - \hat \gamma \hat {\tilde \gamma})$ is the normal Green's function, that is the upper left element in the particle-hole space of Eq.~(\ref{riccati_GF}).

For concreteness we consider the head-to-head DW here. It is convenient to parametrize the spin texture of the effective exchange field by
\begin{eqnarray}
\bm h_{eff} = h_{eff} (\cos \theta, \sin \theta \cos \delta, \sin \theta \sin \delta),
\label{h_texture_hhwall}
\end{eqnarray}
where in general the both angles depend on $x$-coordinate. The equilibrium shape of the DW is dictated by the interplay between the magnetic anisotropy energy and the exchange energy and is given by
 \begin{equation} \label{theta}
 \cos\theta = -  \tanh [(x-x_0)/d_W],
 \end{equation}
 and $\delta = const$ for the classical in-plane DW. 

\section{DOS at the end of the bilayer}

\begin{figure}[!tbh]
   \centerline{\includegraphics[clip=true,width=3.0in]{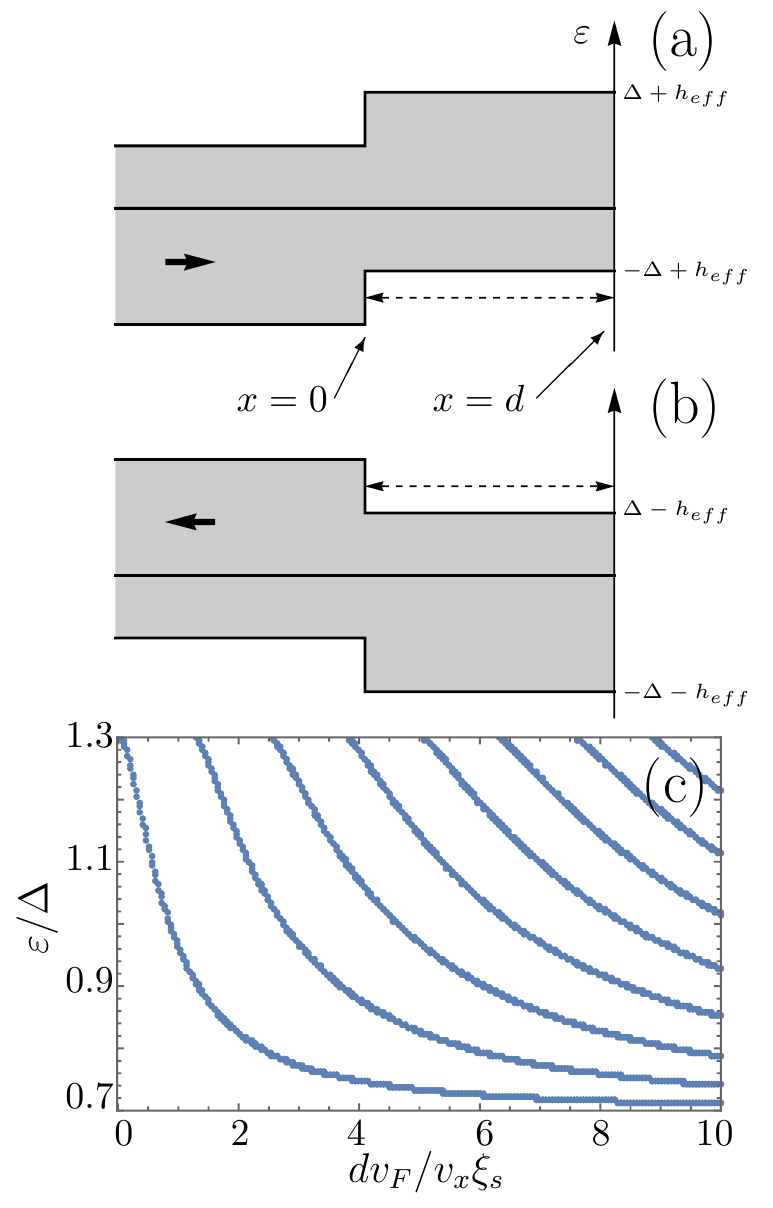}}
     \caption{Schematic picture of the spin-up (a) and spin-down (b) subband DOS. The grey region is a gap for each of the subbands. The dashed lines represent the positions of the bound states (schematic), which are formed as a result of the conventional specular reflection process from the closed end of the superconductor at $x=d$ and the Andreev reflection process from the gap at $x=0$. (c) Bound state energies for spin-down subband as functions of the combination $d/|\cos \alpha|$, where $\alpha$ is the angle, which the quasiparticle momentum makes with the $x$-axis ($h_{eff}=0.3 \Delta$).}
 \label{qualitative}
 \end{figure}

First of all, we consider a toy model of infinitely thin DW  $\bm h_{eff}(x) = -h_{eff}{\rm sign}(x) \bm e_x$. The qualitative picture of the Andreev bound states formation is explained in Figs.~\ref{qualitative}(a)-(b). The bulk DOS for a given spin subband has a conventional BCS-like shape shifted down (up) by $h_{eff}$ for .the spin-up (down) subband. In the region $0<x<d$ the direction of the exchange field is reversed and, consequently, the DOS seen by spin-up quasiparticle is of the bulk spin-down form\cite{note1}.  Therefore, at $-(\Delta+h_{eff})<\varepsilon<-(\Delta-h_{eff})$ the region $0<x<d$ forms a quantum well for spin-up quasiparticles, where the bound states appear as a result of dimensional quantization. For spin-down quasiparticles the same picture is valid at $\Delta-h<\varepsilon<\Delta+h$.  This quantum well can be called "the Andreev quantum well" because the bound state is formed by the combination of the ordinary reflection from the closed end of the superconductor and the Andreev reflection process at $x=0$.

The shortest way to calculate the bound state energies analytically is to solve the Andreev equations for the quasiparticle two-component wave function,  
\begin{eqnarray}
\left(
\begin{array}{cc}
-iv_x \partial_x - \sigma h_{eff}(x) & \Delta \\
\Delta & iv_x \partial_x - \sigma h_{eff}(x)
\end{array}
\right)
\left(
\begin{array}{c}
u_\sigma \\
v_{\bar \sigma} 
\end{array}
\right)
= \nonumber \\
\varepsilon
\left(
\begin{array}{c}
u_\sigma \\
v_{\bar \sigma} 
\end{array}
\right)~~~~~~~~~~~~,
\label{andreev_eq}
\end{eqnarray}
where $\sigma = \pm$ is the spin subband index, $v_x$ is the Fermi velocity component along the $x$-axis and $h_{eff}(x)$ has a step-like form described above. For the more realistic profile of the DW taking into account the finite DW width Eq.~(\ref{andreev_eq}) is not valid because equations for two spin subbands become coupled in this case. Eq.~(\ref{andreev_eq}) should be supplemented by the boundary conditions $(u_\sigma ~~ v_{\bar \sigma})^T(v_x) = (u_\sigma ~~ v_{\bar \sigma})^T(-v_x)$ at the impenetrable surface $x=0$. Solving Eq.~(\ref{andreev_eq}) together with the boundary conditions at the closed end of the superconductor one can find the following equation for the bound state energies:
\begin{equation}
\tan[2\kappa_{1,\sigma}d/|v_x|] = \frac{\kappa_{1,\sigma} \kappa_{2,\sigma}}{\varepsilon^2 - h_{eff}^2 - \Delta^2},
\label{bound_spectrum}    
\end{equation}
where $\kappa_{1,\sigma} = \sqrt{(\varepsilon - \sigma h_{eff})^2-\Delta^2}$ and $\kappa_{2,\sigma} = \sqrt{\Delta^2 - (\varepsilon + \sigma h_{eff})^2}$ are real quantities because the bound states only exist in the range $-(\Delta+h)<\varepsilon<-(\Delta-h)$ for spin-up and in the range $\Delta-h<\varepsilon<\Delta+h$ for spin-down quasiparticles. The solutions of Eq.~(\ref{bound_spectrum}) are represented in Fig.~\ref{qualitative}(c). It is seen that the number of levels grows when $d$ is increased. 

The mechanism of the bound states formation resembles the mechanism of the well-known de Gennes-Saint-James bound states formation due to the inhomogeneity of the superconducting order parameter near the surface (or in the hybrid superconductor/normal metal/insulator structures)\cite{deGennes1963}. However, we would like to note that the physical nature of the bound states discussed here is different because they do not require an order parameter inhomogeneity at the closed end. The order parameter, in principle, should be calculated self-consistently for the problem under consideration, but we do not perform the corresponding calculations here because taking into account the self-consistent form of the order parameter near the closed end does not change the effect qualitatively. Having this physical mechanism of the bound states formation at hand, below we calculate numerically (making use of the method of the Riccati-amplitudes described above) the DOS near the closed end of the superconductor taking into account the realistic profile of the DW magnetization.

\begin{figure}[!tbh]
    \centerline{\includegraphics[clip=true,width=3.0in]{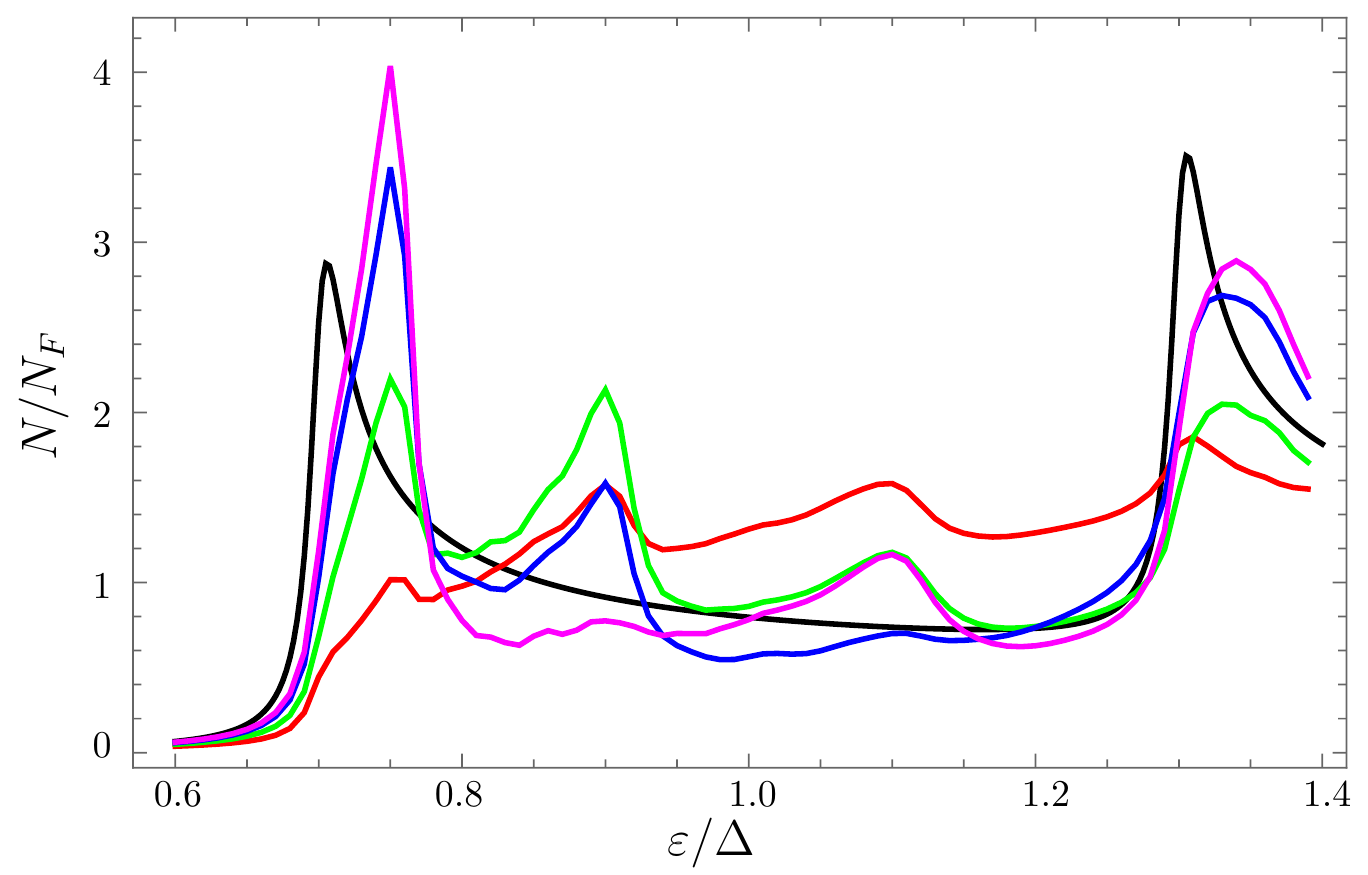}}
     \caption{DOS as a function of quasiparticle energy. Different curves correspond to different locations $x=0$(red), $d/3$(green), $2d/3$(blue), $d$(pink); the black curve is the bulk DOS. $d=3.75 \xi_s$, $d_W=0.25 \xi_s$, $h_{eff}=0.3\Delta$.}
 \label{DOS1}
 \end{figure}

As it was discussed above the bulk DOS of a spin-split superconductor has a typical two-peak shape and is plotted in Fig.~\ref{DOS1} by the black line. The DOS is only shown for positive energies and $N(-\varepsilon)=N(\varepsilon)$.  The other curves in Fig.~\ref{DOS1} demonstrate the DOS at different distances from the closed end of the superconductor for the case when the DW is present in the ferromagnet and its center is at $x=0$. The DOS exhibits several peaks in the energy range between the spin-split coherence peaks. The peaks represent the DOS concentrated at the Andreev bound states, which are localized between the closed end and the DW. The broadening of the peaks is due to the  averaging of the bound state energies over different momenta at the Fermi surface and also due to the finite DW width.  Therefore, in real setups the discussed above Andreev bound states exhibit themselves as smeared Andreev resonances in the momentum-averaged DOS. 

\begin{figure}[!tbh]
    \centerline{\includegraphics[clip=true,width=3.2in]{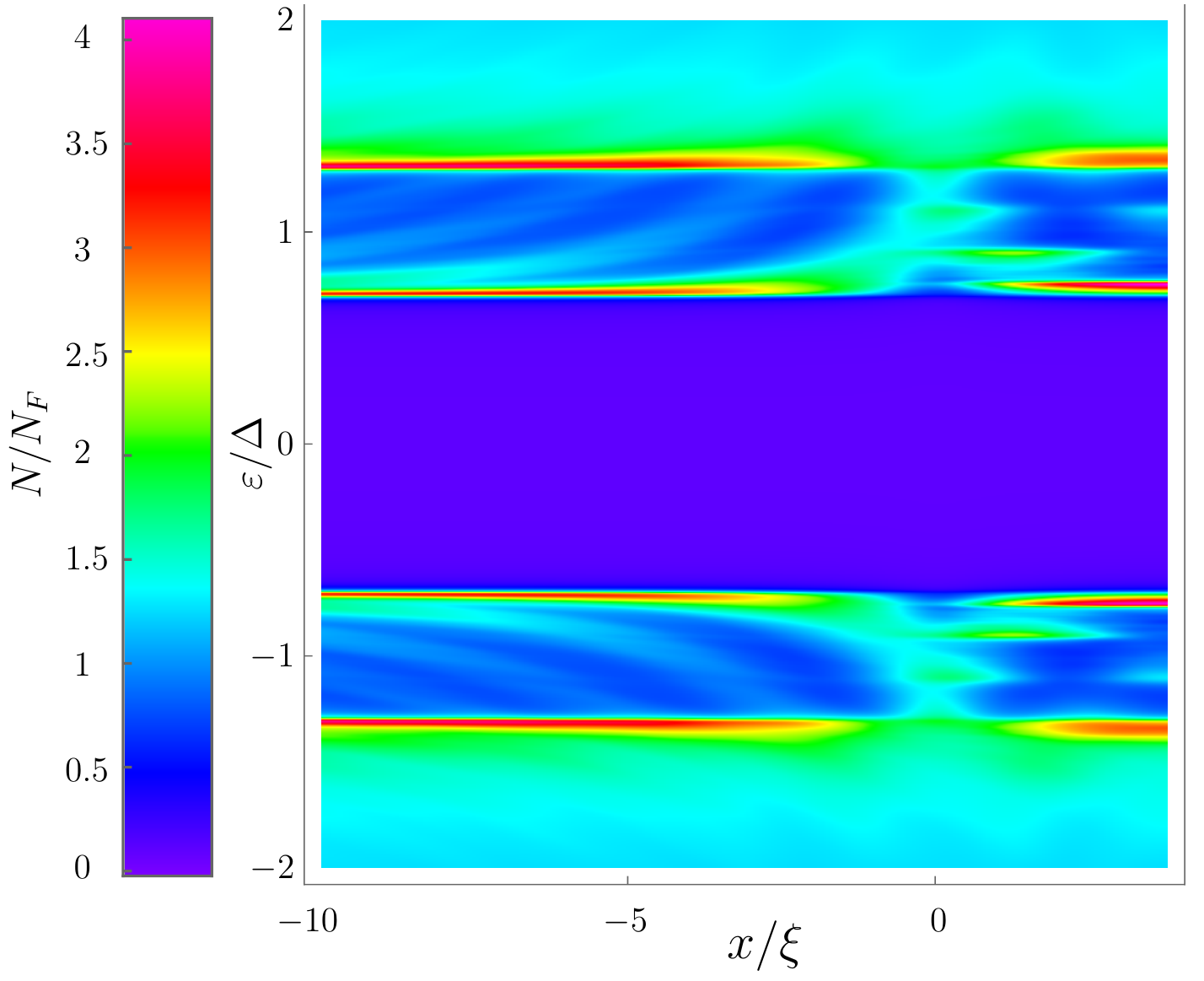}}
     \caption{2D-plot of the DOS in the $(x,\varepsilon)$-plane. The Andreev resonances are seen as horizontal light-green lines at $x>0$ on the sides of the gap region. $d=3.75 \xi_s$, $d_W=0.25 \xi_s$, $h_{eff}=0.3\Delta$.}
 \label{DOS2D}
 \end{figure}

\begin{figure}[!tbh]
    \centerline{\includegraphics[clip=true,width=3.0in]{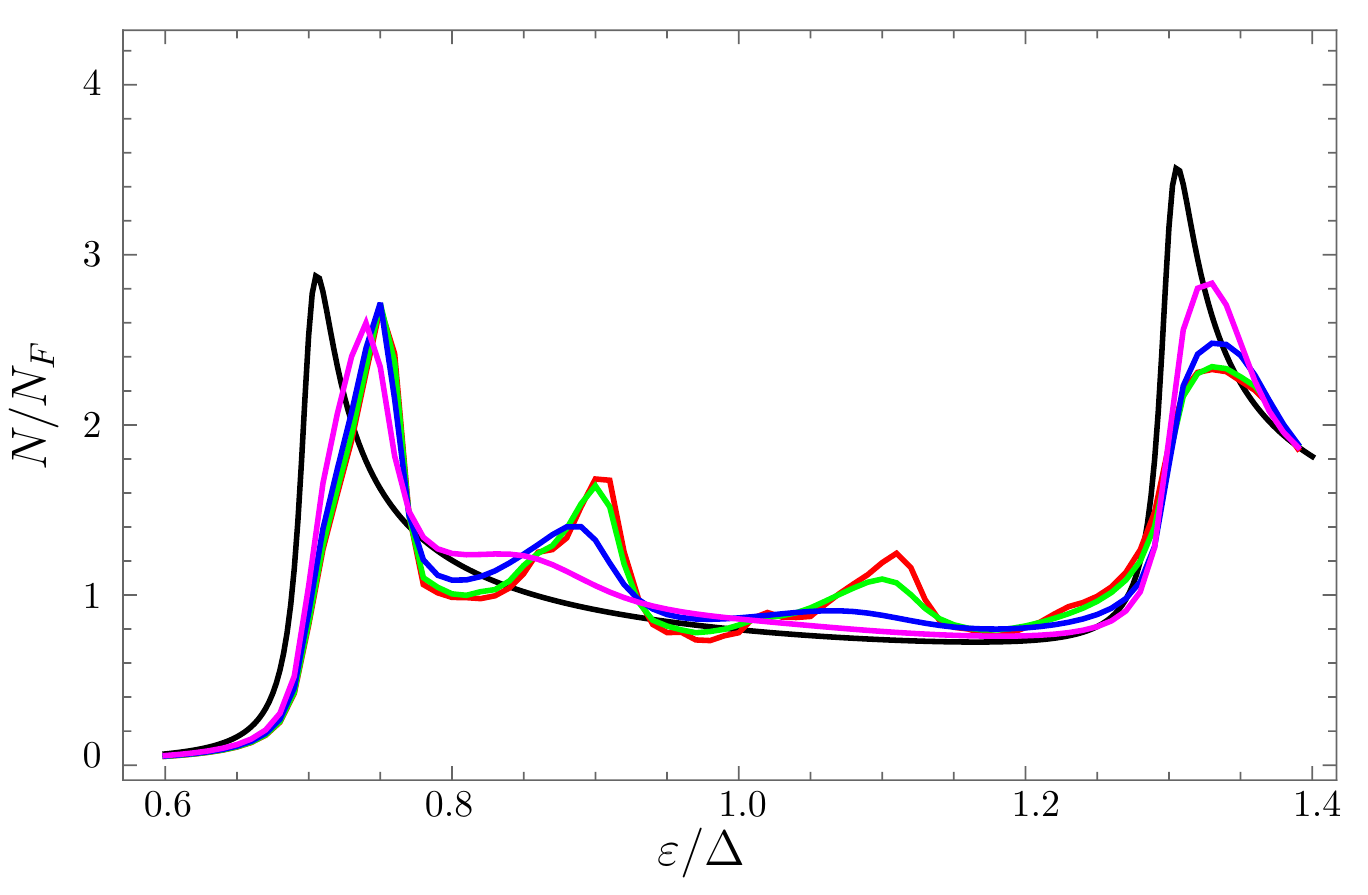}}
     \caption{Coordinate-averaged DOS $\overline N$ for DWs of different width $d_W=0.15\xi_s$(red), $0.25\xi_s$(green), $0.50\xi_s$(blue), $1.0 \xi_s$(pink). The black line is again the bulk DOS. $d=3.75 \xi_s$, $h_{eff}=0.3\Delta$.}
 \label{DOS2}
 \end{figure}

Fig.~\ref{DOS2D} represents a 2D-plot of the DOS in $(\varepsilon,x)$-plane. From Figs.~\ref{DOS1} and Fig.~\ref{DOS2D} it is seen that the peak height at a given peak energy oscillates along the $x$-axis. Therefore, below we study the DOS $\overline{N}$ averaged over the region $0<x<d$. 

\begin{figure}[!tbh]
    \centerline{\includegraphics[clip=true,width=3.0in]{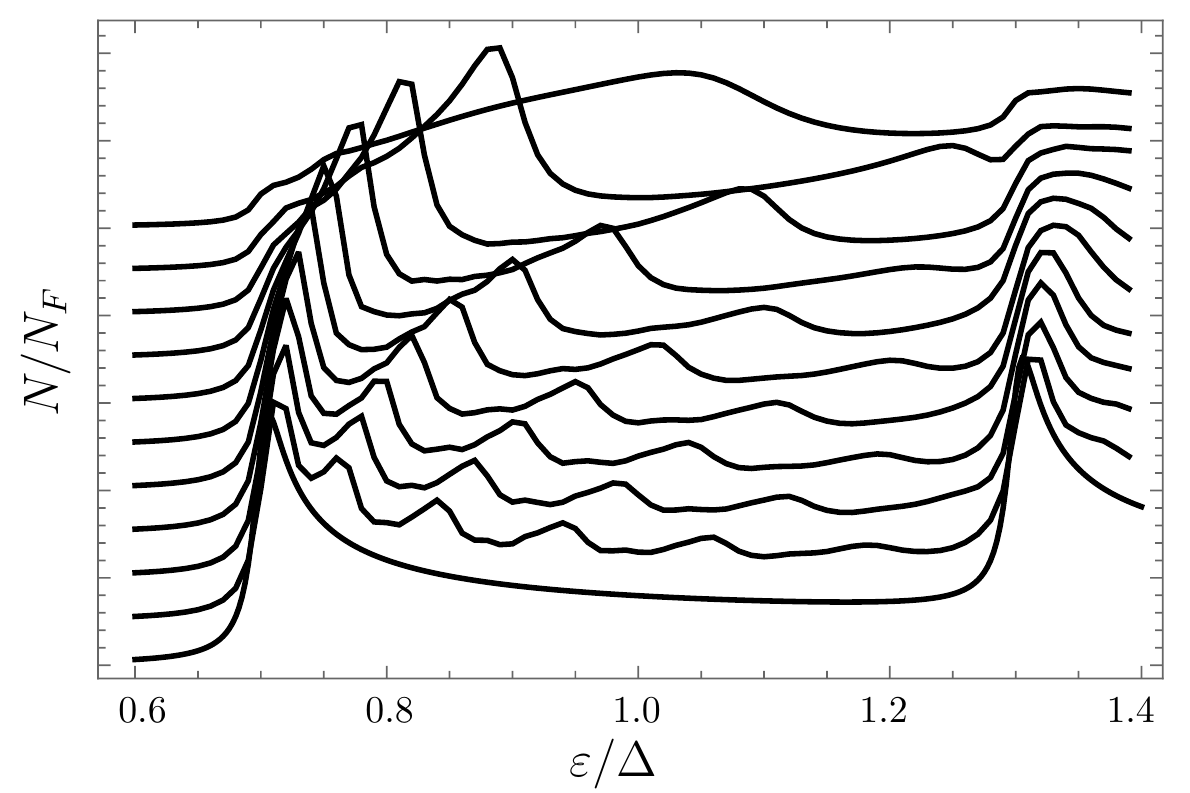}}
     \caption{Coordinate-averaged DOS $\overline N$ for different distances between the DW centre and the  superconductor edge, from bottom to top: $d=\infty,~7.5,~6.75,~6.0,~5.25,~4.5,~3.75,~3.0,~2.25,~1.5,~0.75~\xi_s$. The offset is for clarity. $d_W=0.25 \xi_s$, $h_{eff}=0.3\Delta$.}
 \label{DOS3}
 \end{figure}

$\overline{N}$ for DWs of different widths $d_W$ is presented in Fig.~\ref{DOS2}. It is seen that the resonances smear for wider walls. It is natural because the appearance of the $z$-component of $\bm h_{eff}$ at the DW leads to the nonzero probability for a quasiparticle to reverse its spin at the wall region. Therefore, after the transition through the wall region the spin-up quasiparticle is partially converted to the spin-down one. This spin-down quasiparticle can freely move inside the superconductor according to the illustration in Fig.~\ref{qualitative}. This process provides a leakage channel for the quasiparticles from the quantum well region.  As it should be expected the Andreev resonances for very wide walls practically disappear and the DOS approaches the bulk shape.

Fig.~\ref{DOS3} shows evolution of the DOS when the DW is moved away from the closed end of the superconductor. The Andreev resonances are not pronounced for $d<\xi_S$, upon increasing $d$ more and more resonances appear in the quantum well according to Eq.~(\ref{bound_spectrum}) and Fig.~\ref{qualitative}(c), but their heights get lower and the DOS approaches the bulk form when the DW center goes far from the end of the superconductor. 

\section{conclusions}

The DOS at the end of the superconductor, which is a part of a S/F bilayer, is considered. We study the case when the ferromagnet magnetization has a defect in the form of a DW in the vicinity of the superconductor end. It is shown that in this case the superconducting DOS is reconstructed near the end manifesting spin-split Andreev resonances in the spatial region between the DW and the superconducting end. So far, all types of surface bound states at nonmagnetic surfaces discussed in the literature are known to be formed due to the order parameter absolute value or phase variations  along the quasiparticle trajectory near the surface. On the contrary, these states are formed as a combination of the specular reflection from the closed end and the Andreev reflection from the spin-split gap position shift, which arises due to the DW presence. The order parameter variation along the quasiparticle trajectory near the surface is not required for their arising.

\section{Acknowledgements}
 We thank Wolfgang Belzig for stimulating discussions. 
The work of A.M.Bobkov was supported by the Russian Science Foundation project Nr. 16-42-01050.

\end{document}